\documentclass[preprint,12pt]{elsarticle}

\usepackage{amssymb}

\newcommand{\be}{\begin{equation}}
\newcommand{\ee}{\end{equation}}

\journal{Comptes Rendus Physique}

\begin{document}

\begin{frontmatter}



\title{In social complex systems, the whole can be more or less than (the sum of) the parts}


\author{Eric Bertin}

\address{Univ.~Grenoble Alpes, CNRS, LIPHY, F-38000 Grenoble, France}

\author{Pablo Jensen}

\address{Institut Rh\^one-Alpin des Syst\`emes Complexes, IXXI, F-69342 Lyon, France}

\address{Universit\'e de Lyon, Laboratoire de Physique ENS Lyon and CNRS, 46 Rue d'Italie, F-69342 Lyon, France}

\begin{abstract}
We discuss in a statistical physics framework the idea that ``the whole is less than the parts'', as sometimes advocated by sociologists in view of the intrinsic complexity of humans, and try to reconcile this idea with the statistical physicists wisdom according to which ``the whole is more than the sum of its parts'' due to collective phenomena.
We consider a simple mean-field model of interacting agents having an intrinsic complexity modeled by a large number of internal configurations.
We show by analytically solving the model that interactions between agents lead, in some parameter range, to a `standardization' of agents in the sense that all agents collapse in the same internal state, thereby drastically suppressing their complexity. Slightly generalizing the model, we find that agents standardization may lead to a global order if appropriate interactions are included.
Hence, in this simple model, both agents standardization and collective organization may be viewed as two sides of the same coin.
\end{abstract}

\begin{keyword}



\end{keyword}

\end{frontmatter}


\section{Introduction}

While the polysemy of the word ``complex'' makes any definition of a ``complex system'' somewhat ambiguous, there is general agreement among physicists to define it in the following way \cite{BARRAT}: It is a system composed of a large number of elements interacting without central coordination and spontaneously leading to the emergence of ``complex structures'', i.e., stable structures with patterns that may occur on several spatial or temporal scales. This definition is often summarized by the dictum: ``the whole is more than the sum of the parts'', as these stable structures cannot be deduced from parts in a simple way: 
The viscosity of water is not the average or sum of a hypothetical ``molecular viscosity''.

It should be noted that this standard definition does not take into account any potential specificity of social systems, beyond some trivial specification of the ``parts'' as humans. However, as recognized in the influential review by Castellano et.~al.~\cite{castellano}, there seems to be a radical difference. In the usual applications of statistical physics, ``the macroscopic phenomena are not due to a complex behavior of single entities, but rather to nontrivial collective effects resulting from the interaction of a large number of `simple' elements''.
However, ``humans are exactly the opposite of such simple entities: the detailed behavior of each of them is already the complex outcome of many physiological and psychological processes, still largely unknown''. Castellano et.~al.~justify why physicists crave for finding general models for complex systems, spanning the social/natural realms, using the idea of ``universality'': ``In most situations, qualitative and even some quantitative properties of large-scale phenomena do not depend on the microscopic details of the process [...]. With this concept of universality in mind, one can approach the modelization of social systems, trying \emph{to include only the simplest and most important properties of single individuals} and looking for qualitative features exhibited by models'' \cite{castellano}. 

A first criticism of this homogenization of natural and social systems is political, as it rules out the specifically human capacity to think about and build ``the whole'', instead of letting it emerge ``spontaneously'' \cite{jasss,seuil,Scott,Latour88}. A second criticism draws conclusions from the relative lack of usefulness, in practice, of the idea of universality for analyzing complex social systems. With sociologists, one of us (PJ), has provided another angle for building ``wholes'' from complex social agents, possessing too many internal degrees of freedom to be described as particles with a handful of characteristics.
In this case, what emerges is not larger but ``smaller than the parts'', a kind of ``intersection of individuals'' by standardization of one of their characteristics \cite{whole,JENSEN_CRAS}. This viewpoint is supported by the analysis of digital databases \cite{whole}. Yet, the type of information one gets from database analysis cannot be easily rephrased in the language of statistical physics, which mostly deals with the analysis of collective phenomena through simple models of interacting agents.
It is thus a priori hard to reconcile the physicist's viewpoint according to which ``the whole is more than the sum of its parts'', and the sociologist's viewpoint supporting the idea that the ``the whole is less than its parts'' because humans need to simplify their intrinsic complexity to form a coherent whole in certain circumstances.

In this note, we try to take a first step to reconcile these seemingly opposite viewpoints and discuss a possible meaning of the idea that ``the whole is less than the parts'' in a statistical physics framework.
We first introduce a simple model that can be solved analytically and that provides an illustration of a possible mechanism of standardization of agents driven by their interaction.
In a second step, we slightly generalize the model to allow for the possibility of global order, and find that agents simplification is a prerequisite for global order to appear. Hence, at least in the present toy model, agents standardization and collective organization are intricate phenomena that may be viewed as two sides of the same coin. 
Following the above metaphors, the whole is therefore at the same time \emph{more} and \emph{less} than (the sum of) its parts.


\section{Standardization of agents induced by interactions: a toy model}
\label{sec:model1}

To implement in a minimal way the complexity of agents and the role of interactions, we introduce a simple model composed of $N$ interacting agents with internal states described by a configuration $\mathcal{C} \in \{1,...,H+1\}$. In what follows, we consider that $H$ is a large, but fixed number ---say, e.g., $H=10^4$.
Each agent has a characteric encoded by a variable $S_i(\mathcal{C}) \in \{0,1\}$ (sometimes loosely referred to as a ``spin'' variable in the following, by analogy with statistical physics).
We assume that this characteristic is absent (i.e., $S_i=0$) in most configurations. For the sake of simplicity, we make the hypothesis that the characteristic is present only for a single configuration, say $\mathcal{C}=1$,
so that $S_i(1)=1$ and $S_i(\mathcal{C})=0$ for $\mathcal{C} >1$.
If internal configurations were equiprobable, the value $S_i=1$ would be a rare value unobserved in practice, and one would see only a set of spins $S_i=0$.
Here, we wish to show that interactions between agents can ``standardize'' the agents and make the value $S_i=1$ practically observable.

As for the dynamics, we assume that each agent randomly changes configuration according to a utility function $u_i(\mathcal{C})$ that accounts for interactions with other agents. In other words, agents tend to randomly select configurations that allow them to have more profitable interactions with the other agents.
Inspired by standard models of social interactions (see, e.g., \cite{Nadal04}),
we choose the following simple form for $u_i$:
\be \label{eq:def:utility}
u_i = \frac{K}{N} \sum_{j (\ne i)} S_i S_j \,.
\ee
Starting from a configuration $\mathcal{C}$, a new configuration $\mathcal{C}'$ is picked up with a probability rate
\be \label{eq:transition:rate}
W(\mathcal{C}'|\mathcal{C}) = \frac{1}{1+e^{-\Delta u_i/T}}
\ee
where $\Delta u_i = u_i(\mathcal{C}') - u_i(\mathcal{C})$, the configuration of the other agents being kept fixed.
The parameter $T$ characterizes the degree of stochasticity in the choice of the new configuration, and is analogous to the temperature in statistical physics. For convenience, we call $T$ a temperature in the following, and use the notation $\beta=T^{-1}$.

It is straightforward to show that the variation $\Delta u_i$ of individual utility can be reexpressed as the variation of a global quantity $E$ (the analogue of the global energy in physics, up to a change of sign), $\Delta u_i = \Delta E$, where
\be \label{eq:energy:model1}
E = \frac{K}{2N} \sum_{i,j (i\ne j)} S_i S_j \,.
\ee
In view of the form (\ref{eq:energy:model1}) of the pseudo-energy $E$, the present model shares similarities with both the Ising \cite{LeBellac} and Potts models \cite{review-Potts}, but it actually differs in a significant way from both of these models.

The property $\Delta u_i = \Delta E$ together with Eq.~(\ref{eq:transition:rate}) implies that the dynamics satisfies a detailed balance property with respect to the equilibrium distribution
\be
P(\mathcal{C}_1,\dots,\mathcal{C}_N) \propto e^{E/T} \,.
\ee
To study the behavior of the model, it is convenient to define an order parameter $q$ as
\be
q = \frac{1}{N} \sum_{i=1}^N S_i.
\ee
For large $N$, the quantity $E$ can be rewritten in terms of $q$ as
\be
E = N\varepsilon(q) = \frac{1}{2} N K q^2 \,.
\ee
Note that $E$ is not equal to the total utility $U=\sum_i u_i$; Instead,
one has $E=\frac{1}{2}U$.

\begin{figure}[t]
\centering\includegraphics[width=0.6\linewidth]{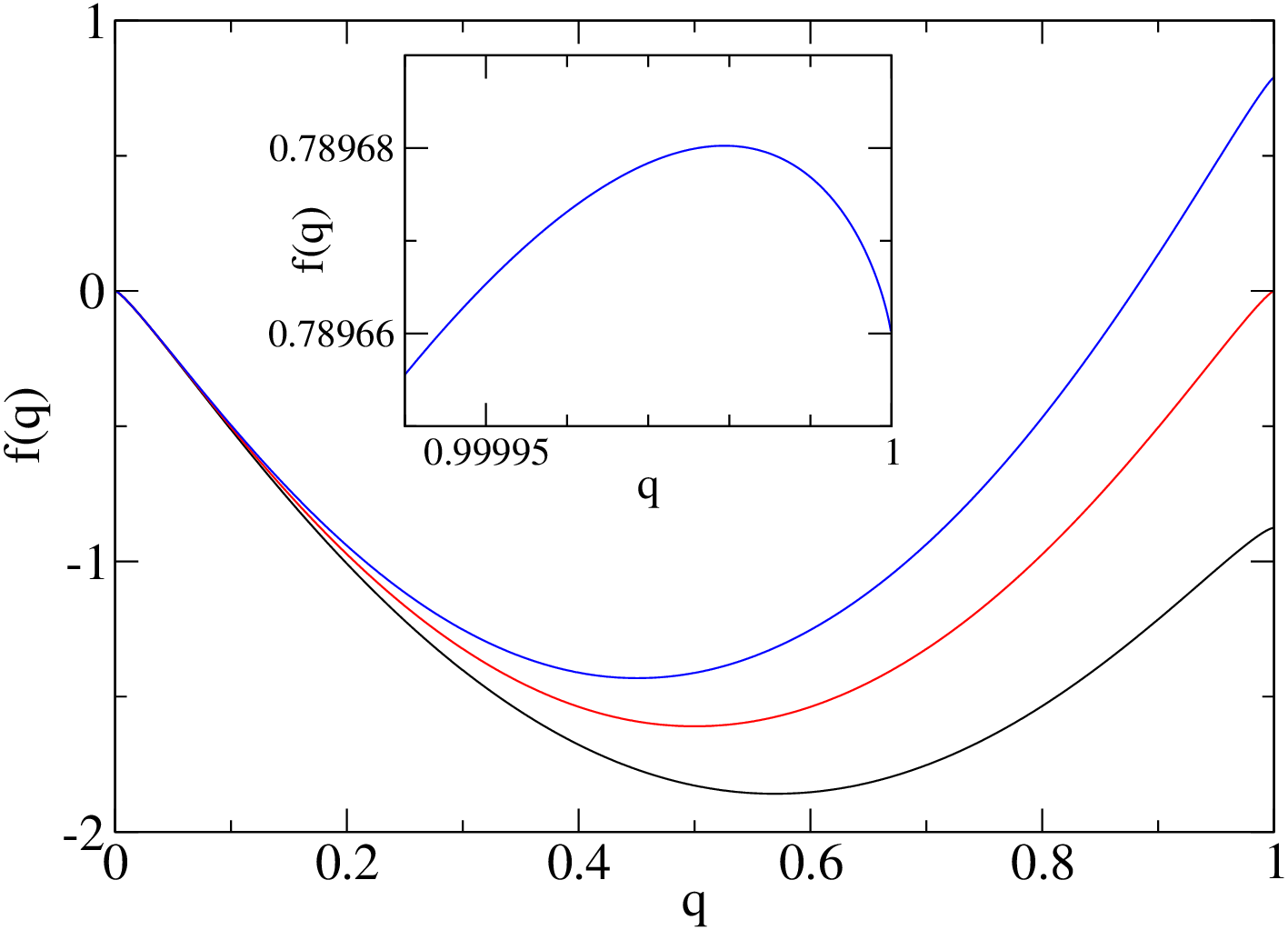}
\caption{Plot of the large deviation function $f(q)$ for $T>T_S$, $T=T_S$ and $T<T_S$ (bottom to top), showing a discontinuous transition in terms of the order parameter $q$. The maximum of $f(q)$ is found at $q \approx 0$ for $T>T_S$ and at $q \approx 1$ for $T < T_S$. Inset: zoom on the curve $T<T_S$, showing the local maximum of $f(q)$ for $q$ very close to $1$. Parameters: $H=10^4$, $K=10$, $T=0.6$, $0.5429$, $0.5$.}
\label{fig:fq}
\end{figure}

We are now interested in the probability distribution $P(q)$, obtained by summing $P(\mathcal{C}_1,\dots,\mathcal{C}_N)$ over all sets of configurations $(\mathcal{C}_1,\dots,\mathcal{C}_N)$ sharing the same value of $q$.
Formally, one has
\be
P(q) = \sum_{\mathcal{C}_1,\dots,\mathcal{C}_N} P(\mathcal{C}_1,\dots,\mathcal{C}_N)
\, \delta \Big( q(\mathcal{C}_1,\dots,\mathcal{C}_N) - q\Big)
\ee
(where $\delta$ stands for a Kronecker delta), leading to
\be
P(q) \propto e^{N[\mathcal{S}(q)+\beta \varepsilon(q)]} \,.
\ee
The entropy $\mathcal{S}(q)$ is given by
\be
\mathcal{S}(q) = \frac{1}{N}\, \ln \Omega(q)
\ee
where $\Omega(q)$ is the number of configurations having a given value of $q$:
\be \label{eq:Omega}
\Omega(q) = \frac{N!}{N_0! N_1!} H^{N_0} ,
\ee
$N_0$ and $N_1$ being the numbers of agents with $S_i=0$ and $1$ respectively.
We thus see that $P(q)$ takes a large deviation form
$P(q) \propto \exp[Nf(q)]$
where the large deviation function $f(q)=\mathcal{S}(q)+\beta \varepsilon(q)$ is independent of $N$, and plays a role similar to the notion of free energy in physics.
Using $q=N_1/N$, one can compute $\mathcal{S}(q)$ from Eq.~(\ref{eq:Omega}), leading for $f(q)$ to
\be
f(q)= -q\ln q -(1-q)\ln (1-q) + \frac{1}{2} \beta K q^2 + (1-q)\ln H \,.
\ee
Fig.~\ref{fig:fq} illustrates the behavior of the function $f(q)$ for different values of temperature, for a given large value of $H$. At high enough temperature $T$, $f(q)$ has a global maximum at $q \approx 0$. Lowering $T$, the global maximum is suddenly shifted to $q \approx 1$ when crossing a characteristic temperature $T_S$.

This behavior is confirmed by a simple analytical calculation, assuming that $H$ is large but finite.
We look for the local maxima of $f(q)$ by computing the derivative $f'(q)$:
\be
f'(q) = -\ln q + \ln (1-q) +\beta K q -\ln H =0 \,.
\ee
Looking first for a solution $q_1 \ll 1$, we find $q_1 \approx 1/H$. 
At low enough temperature $T$, another solution $q_2 \approx 1-H e^{-K/T}$ also exists.
It can be checked that $q_1$ and $q_2$ are local maxima.
Note that a local minima also exists between $q_1$ and $q_2$.
To check the relative stability of $q_1$ and $q_2$, we compute $f(q_1)$ and $f(q_2)$. 
Defining $T_S$ as
\be \label{eq:def:Ts}
T_S = \frac{K}{2\ln H} \,,
\ee
we find that $f(q_1)>f(q_2)$ for $T>T_S$, and $f(q_1)<f(q_2)$ for $T < T_S$
(see Fig.~\ref{fig:fq}).
Thus $q_1 \approx 0$ is the globally stable state for $T>T_S$ and $q_2 \approx 1$ is the globally stable state for $T < T_S$.
The transition between these two states is therefore discontinuous in terms of the order parameter $q$, which essentially jumps from $0$ to $1$.

This means that for strong enough coupling, or low enough temperature $T$, agents become ``standardized'' and have a non-zero value of their spin $S_i$, in the sense that they spend most of their time in the state $S_i=1$.
In other words, thanks to their interaction, the complexity of agents has been suppressed in the low temperature regime, and agents could be modeled in this context as simple and stable ``social atoms'' \cite{JENSEN_CRAS} that have the permanent characteristic corresponding to $S_i=1$.

Clearly, the complexity of agents remains minimal in this toy model, as it is encoded only in the fact that agents have many internal configurations.
One may guess, though, that the above standardization transition could still be present if for instance all configurations with $S_i=0$ were not equivalent, and had a more complicated dynamics, provided that the interaction term present in the utility (\ref{eq:def:utility}) remains the dominant one.
Further work would be needed to confirm this expectation.

\section{Emergence of global order out of agents standardization}
\label{sec:model2}

Having seen how an agent characteristic (or ``spin'') could emerge from interactions between agents, it is tempting to slightly generalize the model in order to introduce the potentiality of a symmetry breaking, that would mimic in a minimal way the emergence of some form of organization (``order'') in a social group.
With this goal in mind, we now consider a variant of the above model,
where the variables $S_i$ now take three values,
$S_i(\mathcal{C}) \in \{-1,0,1\}$, where now $\mathcal{C} \in \{1,\dots,H+2\}$.
We assume for simplicity that $S_i(1)=1$, $S_i(2)=-1$ and $S_i(\mathcal{C})=0$ for $\mathcal{C}>2$.
The emergence of a specific characteristic at the agent level would now result from the interactions of the variables $|S_i|$ between agents, while the emergence of a global order would result from the interactions of the (nonzero) $S_i$ variables.
We thus choose as a simple form of the utility function
\be \label{eq:def:utility2}
u_i = \frac{K}{N} \sum_{j (\ne i)} |S_i|\, | S_j| + \frac{J}{N} \sum_{j (\ne i)} S_i S_j \,.
\ee
The dynamics of internal configurations of agents is still defined by the transition rate given in Eq.~(\ref{eq:transition:rate}),
with the utility $u_i$ now given by Eq.~(\ref{eq:def:utility2}).
The stochastic dynamics satisfies a detailed balance with respect to the 
equilibrium distribution $P(\mathcal{C}) \propto e^{E/T}$, where $E$ is given by
\be
E = \frac{K}{2N} \sum_{i,j (i\ne j)} |S_i|\, |S_j| + \frac{J}{2N} \sum_{i,j (i\ne j)} S_i S_j \,.
\ee
To characterize both the possible standardization of agents and the possible onset of order, we introduce two order parameters $q$ and $m$, defined as
\be
q = \frac{1}{N} \sum_{i=1}^N |S_i|, \quad
m = \frac{1}{N} \sum_{i=1}^N S_i \,.
\ee
The order parameter $q$ is the same as the one of the model studied in Sect.~\ref{sec:model1}, and $m$ plays the same role as the magnetization in the Ising model for instance (see, e.g., \cite{LeBellac}).
The pseudo-energy $E$ satisfying $\Delta u_i = \Delta E$ can then be rewritten for large $N$ as
\be
E = N\varepsilon(q,m) =  \frac{1}{2}N K q^2 + \frac{1}{2}N J m^2 \,.
\ee
From this point of view, the present model may thus be considered as a generalization of the spin-1 model introduced by Blume, Emery and Griffiths \cite{BEG71}.
Note that $m$ and $q$ satisfy the constraint $|m| \le q$.
In the same way as above, we determine the joint distribution $P(q,m)$ of the two order parameters:
\be \label{eq:Pqm}
P(q,m) \propto e^{N[\mathcal{S}(q,m)+\beta \varepsilon(q,m)]}
\ee
with $\mathcal{S}(q,m)$ the entropy. To evaluate the latter, we first determine the number $\Omega(q,m)$ of configurations having a given value of $q$:
\be
\Omega(q,m) = \frac{N!}{N_{-1}! N_0! N_1!} H^{N_0}
\ee
where $N_{-1}$, $N_0$ and $N_1$ are the numbers of spins $S_i=-1$, $0$ and $1$ respectively. Using $q=1-N_0/N$ and $m=(N_1-N_{-1})/N$, one can compute the entropy
\be
\mathcal{S}(q,m) = \frac{1}{N} \, \ln \Omega(q,m),
\ee
as well as the large deviation function $f(q,m)=\mathcal{S}(q,m)+\beta \varepsilon(q,m)$, defined from Eq.~(\ref{eq:Pqm}).
One finds the explicit result
\begin{eqnarray} \nonumber
f(q,m) &=& - (1-q)\ln(1-q) - \frac{1}{2}(q-m)\ln(q-m)- \frac{1}{2}(q+m)\ln(q+m)\\
&& \qquad\qquad + \frac{1}{2}\beta Kq^2 + \frac{1}{2}\beta Jm^2 + (1-q) \ln H + q \ln 2 \,.
\end{eqnarray}
The local maxima are determined by solving the equations
$\partial f/\partial q = \partial f/\partial m = 0$, with $|m| \le q$.
The analysis is slightly more involved than in the previous case, but results can be summarized as follows.
To make the main message clearer, we first leave aside one technical subtlety, and come back to it afterwards.
Let us define the two characteristic temperatures
\be \label{def:TS:TM}
T_S = \frac{K}{2\ln \frac{H}{2}} \,, \quad T_M = J \,.
\ee
The expression of $T_S$ is the same as the expression of $T_S$ found in Eq.~(\ref{eq:def:Ts}), except that $H$ has been replaced by $\frac{H}{2}$ (to be interpreted as the ratio of the number of configurations with $S_i=0$ over the number of configurations with $|S_i|=1$).
For $T > T_S$, the state ($m_1=0, q_1 \approx 0$) is globally stable (we recall that $H$ is assumed to be large).
For $T < T_S$, the stable state corresponds to $q_1 \approx 1$, and the value of $m$ depends on the respective values of $T_S$ and $T_M$.

\begin{figure}[!ht]
\centering\includegraphics[width=0.6\linewidth,clip]{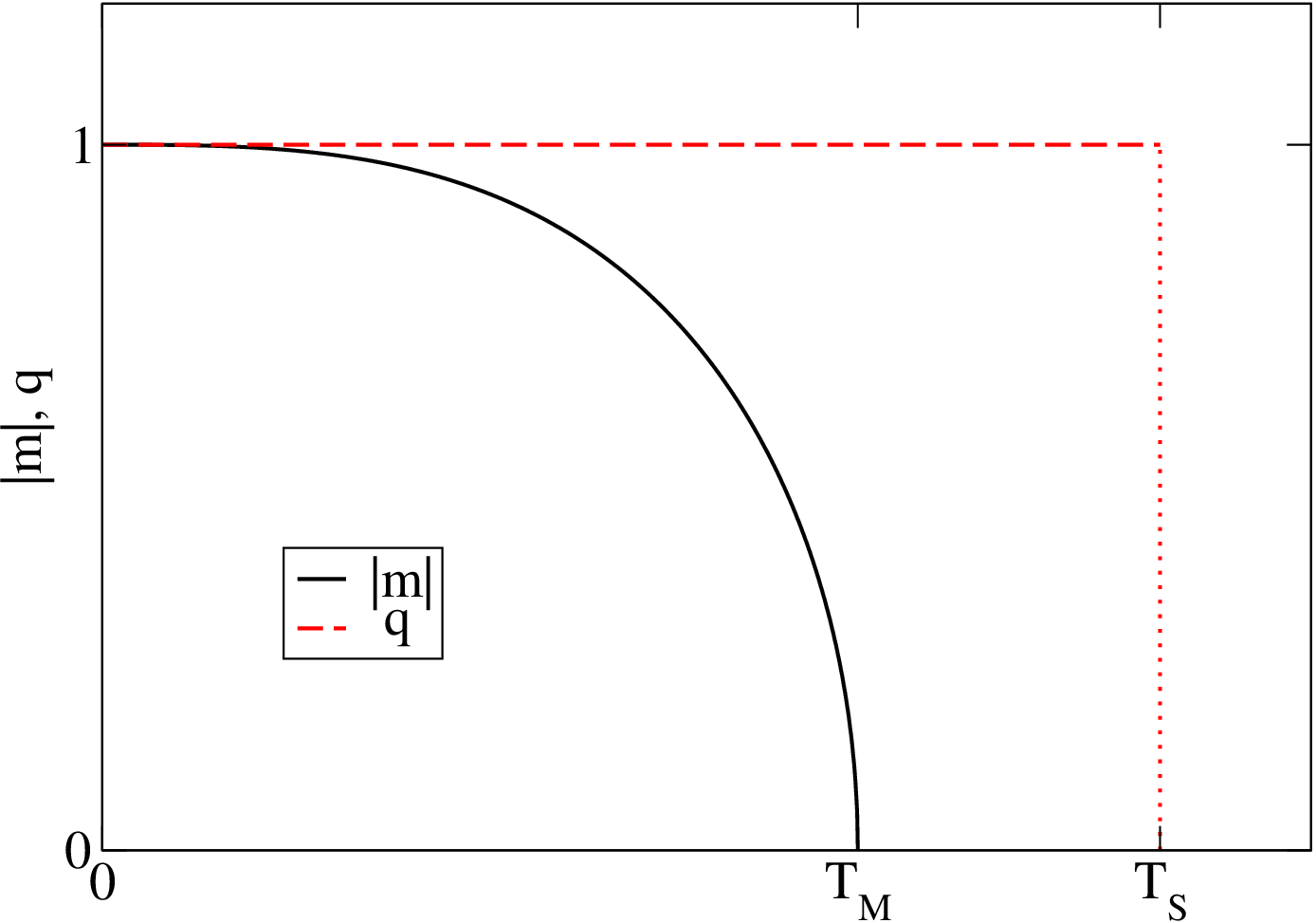}

\medskip
\centering\includegraphics[width=0.6\linewidth,clip]{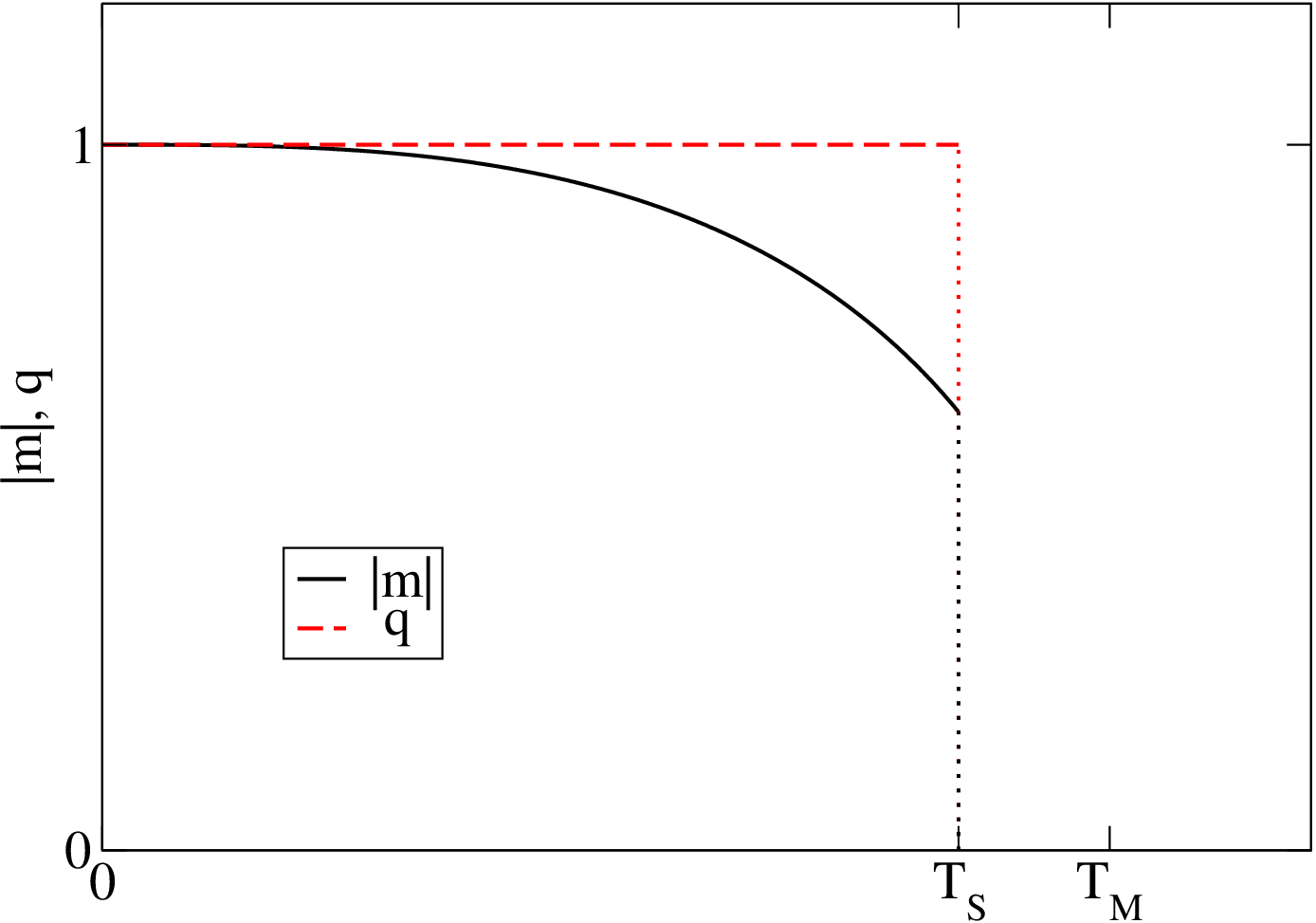}
\caption{Schematic phase diagram showing the behavior of the two order parameters $q$ and $m$ as a function of temperature $T$. Top: case $T_M < T_S$; when lowering temperature, agents first `standardize' at $T=T_S$ ($q$ jumps from $0$ to $1$), and then global order appears at the lower temperature $T_M$ ($|m|$ continuously increases from $0$).
Bottom: case $T_S < T_M$; nothing happens at $T_M$, both order parameters $q$ and $|m|$ jump to a nonzero value at the lower temperature $T_S$. Hence both agent standardization and onset of order appear in a discontinuous way, at the same temperature $T_S$.}
\label{fig:phasediag}
\end{figure}

If $T_M < T_S$, a non-zero magnetization appears only when $T<T_M$, and the onset of order is continuous. One thus has for $T_M < T < T_S$ that $q \approx 1$ (agents are standardized), but $m=0$ (no spontaneous symmetry breaking).
When $T<T_M$, the magnetization takes a non-zero value $m=\pm m_0$, with $m_0$ given for $T$ close to $T_M$ by
\be
m_0 \approx \sqrt{3\left(\frac{T_M-T}{T}\right)}.
\ee

Instead, if $T_S < T_M$, the ``magnetization'' $m$ jumps to a finite value $\pm m_S$ when lowering $T$ below $T_S$, as soon as $q$ jumps to $1$.
There is no direct signature of $T_M$ in the behaviour of the order parameters.
The value of $m_S$ can be approximated in two limits.
If $T_M \gg T_S$, $m_S \approx 1$, while if $T_M-T_S \ll T_S$, one has
\be \label{eq:mS}
m_S \approx \sqrt{3\left(\frac{T_M-T_S}{T_S}\right)}.
\ee
The schematic phase diagrams of the model in the two cases $T_M < T_S$ and $T_S < T_M$ are displayed in Fig.~\ref{fig:phasediag}.

We now come back to the technical subtlety alluded to above.
When $T_S<T_M$ according to the definitions (\ref{def:TS:TM}), a careful look at the calculation actually shows that the emergence of a non-zero magnetization slightly shifts the value of $T_S$ to larger values, so that $T_S$ should actually read (still in the limit of large $H$)
\be \label{def:TSnew}
T_S = \frac{K+Jm_S^2}{2\ln \frac{H}{2}},
\ee
where $m_S$ is self-consistently determined by Eq.~(\ref{eq:mS}).
When $T_S$ and $T_M$ are close, one has $J \approx K/[2\ln(H/2)] \ll K$ as well as $m_S^2 \ll 1$, so that the correction to $T_S$ is actually small with respect to the value given in Eq.~(\ref{def:TS:TM}).
But if $T_S \ll T_M$, $m_S \approx 1$ and $J$ is not necessarily small with respect to $K$ so that the correction to $T_S$ may be significant. In any case, this shift in the value of $T_S$ does not change the picture presented in Fig.~\ref{fig:phasediag}.

To sum up, when $T_M < T_S$, the emergence of order below $T_M$ is simply the usual transition of the Ising model, so that simplifying the agents to simple Ising spins $S_i = \pm 1$ turns out to be legitimate in this case, if one is interested only in modeling the onset of global order associated with a symmetry breaking.
In contrast, when $T_S < T_M$, this simplification is no longer legitimate, and the emergence of order appears only below $T_S$, once agents are ``standardized''. 

Coming back to the case $T_M < T_S$ one may intuitively describe the standardization occuring at $T_S$ as ``the whole is less than its parts'' because agents leave apart their complexity, while the transition at $T_M$ corresponds to the more usual (in physics) transition to order interpreted as ``the whole is more than the sum of its parts'', because of the emergence of collective order.

\section{Discussion}

We have seen in the above toy models how the point of view that complex agents need to reduce their complexity to form a coherent group can to some extent be reconciled with the idea that simple agents may form collective patterns through their interactions.
To go beyond the above toy models and try to better grasp the complexity of agents, one may consider several types of variables $S_i^{(k)}(\mathcal{C})$ characterizing different characteristics of the same agent. Interactions between agents could then favor the emergence of one characteristic or another, in turn allowing for the onset of some type of collective organization if enough agents have standardized with the same characteristic.
Some type of disorder could also be included so that all the internal configurations of agents have different properties.

In addition, it would of course be of interest to study the effect of connectivity of the social network. We have here considered the simplest, fully-connected geometry where all agents interact in the same way with all other agents. Such a connectivity might be relevant in the case of a small group (for instance a music band \cite{JENSEN_CRAS}), but surely cannot account for the intricate connectivity of larger societies, where people are connected through different, and sometimes entangled, communities.
Further work would be needed to investigate the impact of a lower connectivity on the transitions reported here, especially on the ``standardization transition'' of individual agents ---the symmetry breaking transition being of course known to depend on connectivity (see, e.g., \cite{LeBellac,Chaikin}).


\bibliographystyle{elsarticle-num-names} 
\bibliography{biblio-cras.bib}

\end{document}